\documentclass[epj,pacs]{svjour}

\usepackage{dcolumn}
\usepackage{graphicx}
\usepackage{epsf}
\usepackage{amsmath}
\usepackage{bm}
\usepackage{bbm}
\usepackage{color}
\usepackage{cite}
\usepackage{amsmath}
\usepackage{amssymb}
\usepackage{revsymb4-1}

\newcommand{\ii}{\mathrm i}
\newcommand{\fm}{\mathrm{fm}}

\newcommand{\ato}{\mathrm{ATP}}
\newcommand{\nuc}{\mathrm{FMS}}

\journalname{EPJ D}

\begin{document}

\setcounter{page}{1}

\sloppy

\title{Proton Radius, Darwin--Foldy Term and Radiative Corrections}

\author{U. D. Jentschura}

\institute{Department of Physics, Missouri University of Science
and Technology, Rolla MO65409, USA \and
Institut f\"ur Theoretische Physik,
Universit\"{a}t Heidelberg,
Philosophenweg 16, 69120 Heidelberg, Germany}

\date{Received: 2010}

\abstract{We discuss the role of the so-called Darwin-Foldy term in the evaluation
of the proton and deuteron charge radii from atomic hydrogen spectroscopy
and nuclear scattering data. The question of whether this 
term should be included or excluded from the nuclear radius 
has been controversially discussed in the literature. We attempt to
clarify which literature values correspond to which conventions.
A detailed discussion of the conventions appears useful because a 
recent experiment [R. Pohl {\em et al.}, Nature {\bf 466}, 213 (2010)]
has indicated that there is a discrepancy between the proton 
charge radii inferred from ordinary (``electronic'') 
atomic hydrogen and muonic hydrogen.
We also investigate the role of quantum
electrodynamic radiative corrections in the determination of nuclear radii from
scattering data, and propose a definition of the nuclear self energy which is
compatible with the subtraction of the radiative corrections in scattering
experiments.}

\maketitle

%
%
\section{Introduction}
\label{intro}

Quite surprisingly, 
the recent muonic hydrogen Lamb shift experiment~\cite{PoEtAl2010}
at PSI has led to a value of the proton charge radius 
which is in disagreement with both the 2006 CODATA value of the 
mean-square proton charge radius~\cite{MoTaNe2008}, as well as in disagreement with the 
mean-square charge radius derived from electron-proton 
scattering experiments~\cite{Ro2000proton,Si2003proton,Si2007}. 
This disagreement raises a number of questions, 
two of which are the following.

{\em (i)} Is it possible that different conventions have been used
in order to infer the mean-square charge radius of the proton 
in atomic and nuclear physics? In particular, in 
Ref.~\cite{FrMaSp1997}, the authors advocate to add the 
so-called Darwin--Foldy correction to the proton charge 
radius. 
Yet, its inclusion either into the
nuclear radius~\cite{FrMaSp1997} or into the electron binding
energy~\cite{PaKa1995} has been the subject of discussions and is known to
depend on the spin of the nucleus~\cite{PaKa1995}.
The question is whether the inclusion or exclusion of the 
Darwin--Foldy correction has been implemented consistently
in all determinations of the proton charge radius in atomic  
and nuclear physics.

{\em (ii)} The quantum electrodynamic (QED) radiative corrections
to electron-proton scattering have been discussed in a 
number of papers, notably, 
Refs.~\cite{MoTs1969,MaTj2000,VaEtAl2000,EnEtAl2001,BlMeTj2005}.
The QED corrections are subtracted before the form factors
of the proton are deduced from experiment. The notion
is that the electric and magnetic Sachs form factors of the 
proton ($G_E$ and $G_M$) should be defined so that they 
correspond to the internal structure of the proton.
The same applies to the mean-square proton charge radius, 
which is proportional to the slope of the $G_E$ form factor 
at zero momentum transfer.
The corresponding correction in atomic physics is the 
so-called nuclear self-energy~\cite{Pa1995radrec}. 
The question is whether the subtraction of the 
radiative corrections in scattering experiments 
are compatible with the common definition of the 
nuclear self-energy used in the atomic physics literature.

Here, we attempt to answer both of these questions, and we
also investigate the role of QED radiative corrections in the 
determination of the nuclear radius.
Strictly speaking, the slope of the Dirac form factor
$F_1$ and of the Sachs form factor $G_E$ of the proton is known to be
infrared divergent for any spin of the nucleus~\cite{ItZu1980},
unless quantum electrodynamic (QED) radiative corrections are subtracted.
From the atomic physics point of view, this divergence is
manifest in a logarithmic term in the nuclear self-energy
which contributes to the atomic binding energy.
Radiative corrections and infrared bremsstrahlung effects
(which depend on the acceptance range of the detectors)
are subtracted before evaluating
the slope of the form factors. As a cultural matter, this aspect is 
not mentioned in the pertinent literature~\cite{Ro2000proton,Si2003proton,Si2007}.

From scattering experiments, we have a rather
old value from Ref.~\cite{BoPeSiWaWe1974} for the 
root-mean-square proton charge radius $r_p = \sqrt{ \langle r^2 \rangle_p}$,
which reads $r_p = 0.88(3) \, \fm$. 
It was obtained using a dipole fit to the form factor.
In Ref.~\cite{Ro2000proton}, this value has been
confirmed and improved to yield $r_p = 0.880(15) \, \fm$.
Then, according to Refs.~\cite{Si2003proton,Si2007},
the proton radius inferred from the world scattering data reads
\begin{equation}
\label{rmsproton}
r_p = 0.895(18) \, \fm \,,
\end{equation}
in good agreement with the 2008 CODATA value of
\begin{equation}
\label{rmspCODATA}
r_p = 0.8768(69) \, \fm \,.
\end{equation}
The latter value is mainly inferred from the analysis of spectroscopic
data from atomic hydrogen and deuterium spectroscopy~\cite{JeKoLBMoTa2005,MoTaNe2008}.
The most recent and accurate measurement of the proton radius
from electron scattering~\cite{BeEtAl2010}, yields a value of
\begin{equation}
\label{rpSCAT}
r_p = 0.879(8) \, {\rm fm}  \,,
\end{equation}
when the statistical and systematic uncertainties given in
Ref.~\cite{BeEtAl2010} are added quadratically, in excellent 
agreement with the CODATA value~\eqref{rmspCODATA} 
inferred mainly from atomic spectroscopy and the value~\eqref{rmsproton}
inferred from the world average of scattering data.
However, there is a large discrepancy with the 
recent value from the PSI muonic hydrogen 
experiment, which reads
\begin{equation}
\label{rpMUONIC}
r_p = 0.84184(67) \, {\rm fm}  \,.
\end{equation}
Because of this discrepancy, a study of the conventions 
used in the determination of the proton radius 
from scattering data and spectroscopy is indicated.

We proceed as follows. First, the role of the Darwin-Foldy correction
in atomic and nuclear physics determinations of the 
proton charge radius is analyzed (Sec.~\ref{dfterm}).
Radiative corrections to the proton line are discussed in Sec.~\ref{radcorrec}.
Conclusions are reserved for Sec.~\ref{conclu}.
SI (MKSA) units are used throughout the article unless stated otherwise.

%
%
\section{Darwin--Foldy Correction}
\label{dfterm}

In Ref.~\cite{PaKa1995}, it has been shown that the zitterbewegung term
of the nucleus is absent in the atomic binding energy 
for spin-$0$ and spin-$1$ nuclei such as the deuteron.
We here use the conventions of Ref.~\cite{PaKa1995},
acknowledging that others exist~\cite{KhMiSe1996}.
From the nuclear physics side~\cite{FrMaSp1997}, 
it has been recommended to include the so-called Darwin--Foldy 
correction in the value of the proton radius. 
There is a connection between these two statements which will
be explored in the following.

The zitterbewegung term of the nucleus is part of the 
so-called Barker-Glover corrections to atomic energy levels~\cite{BaGl1955}.
The Barker-Glover corrections follow from the two-body Breit Hamiltonian 
(Chap.~83 of Ref.~\cite{BeLiPi1982vol4}) and are listed in the 
last term on the right-hand side of Eq.~(10) of Ref.~\cite{MoTaNe2008}.
They read
\begin{equation}
E_{\rm BG} = \frac{(Z\alpha)^4 m_r^3 \, c^2}{2 n^3 m_N^2} 
\left( \frac{1}{j+ 1/2} - \frac{1}{\ell+1/2} \right) \,
(1 - \delta_{\ell 0}) \,,
\end{equation}
where $Z$ is the nuclear charge number, 
$\alpha$ is the fine-structure constant, 
$m_r$ is the reduced mass of the system, 
$m_N$ is the nuclear mass,
$c$ is the speed of light, $n$ the main quantum number,
and $j$ and $\ell$ are the total and the orbital angular momentum 
quantum numbers of the reference state.
Indeed, the last term in this expression is the Darwin--Foldy term,
\begin{align}
E_{\rm DF} = & \;
-\frac{(Z\alpha)^4 m_r^3 c^2}{2 n^3 m_N^2} 
\left( \frac{1}{j+ 1/2} - \frac{1}{\ell+1/2} \right) \, 
\delta_{\ell 0} 
\nonumber\\[0.5ex]
=& \; \frac{(Z\alpha)^4 m_r^3 c^2}{2 n^3 m_N^2} \, \delta_{\ell 0} \,,
\end{align}
which is due to the zitterbewegung term of the nucleus.
Alternatively, it can be written as
\begin{equation}
\label{EDF}
E_{\rm DF} = \frac23 \; 
\left( \frac{m_r}{m_e} \right)^3 \; 
\frac{(Z\alpha)^4 m_e c^2}{n^3\,\lambdabar_C^2} \; 
\left\{ \frac{3 \hbar^2}{4 m_N^2 c^2} \right\}\, 
\delta_{\ell 0} \, .
\end{equation}
Here, $\alpha$ is the fine-structure constant,
and $Z$ is the nuclear charge number,
$m_r$ is the reduced mass of the system,
$m_e$ is the electron mass,
$c$ is the speed of light,
and $\lambdabar_C$ is the Compton wavelength of the electron divided by
a factor $2 \pi$. The Darwin-Foldy term is nonvanishing only for $S$ 
states ($\ell = 0$).

The main nuclear-size shift of atomic energy levels is given by 
the well-known expression 
\begin{equation}
\label{deltaENS}
E_{\textrm{NS}} = 
\frac23 \left( \frac{m_r}{m_e} \right)^3 \,
\frac{(Z \alpha)^4 m_e c^2}{n^3 \, \lambdabar_C^2} \, 
\langle r^2 \rangle_N \,
\delta_{\ell 0} \,,
\end{equation}
where $\langle r^2 \rangle_N$ is the mean-square charge radius of the 
nucleus. The expressions~\eqref{EDF} and~\eqref{deltaENS} have the 
same structure. 
The nuclear size effect and the Darwin--Foldy correction have the 
same effect on the spectrum if we alternatively add the 
Darwin--Foldy correction to the atomic energy levels or if we add
the term in curly brackets in Eq.~\eqref{EDF} to the mean square nuclear radius.
Let us refer to the nuclear radius defined without the 
Darwin-Foldy term as the nuclear radius in ``atomic physics'' (ATP) conventions.
Indeed, this definition has implicitly been proposed in the 
paper~\cite{HuEtAl1998}, where the nuclear radius difference of proton 
and deuteron was experimentally measured and theoretically evaluated.
The following two replacements are found to be equivalent,
\begin{equation}
\label{repl}
E_{\rm NS} \to E_{\rm NS} + E_{\rm DF} \;
\Leftrightarrow \;
\left< r^2 \right>^\ato_N \to 
\left< r^2 \right>^\ato_N +
\left< r^2 \right>^{\rm DF}_N  \,,
\end{equation}
where
\begin{equation}
\left< r^2 \right>^{\rm DF}_N =
\frac{3 \hbar^2}{4 m_N^2 c^2} \,.
\end{equation}
Here, $N$ denotes the nucleus ($N = p$ for the proton),
and $m_N$ is the nuclear mass. According to 
Ref.~\cite{FrMaSp1997}, in atomic physics
conventions, the proton charge radius is proportional to 
the slope of a subtracted Sachs form factor $G_E$,
\begin{equation}
\label{rATP}
\langle r^2 \rangle^\ato_p =
\langle r^2 \rangle^p_E = 
6 \hbar^2 \left. 
\frac{\partial G_E(q^2)}{\partial q^2} \right|_{q^2 = 0} \,,
\end{equation}
where $q^2 = (q^0)^2 - \vec q^2$ is the momentum transfer.
In order to be consistent, 
it is important to stress that a subtracted
form factor $G_E$ has to be used
in the evaluation of the slope, because formally,
the Sachs form factor $G_E$ has an infinite slope at zero 
momentum transfer, due to an infrared  divergence at the 
one-loop level, which is caused by vertex corrections
involving virtual interactions with very soft virtual photons. 
This is illustrated in the discussion following Eq.~\eqref{separation}
below. The infrared divergence thus is of quantum electrodynamic origin and 
not a consequence of the internal structure of the proton,
and it only enters at the one-loop level (order $\alpha$).
In the literature, the subtractions are sometimes carried out tacitly,
and it is understood that the form factor $G_E$ 
employed for the proton
is due entirely to its internal structure, and all QED effects have
been subtracted.

Let us set this problem aside for the moment and continue to study 
the role of the Darwin--Foldy correction in the definition of the 
nuclear charge radius.
An alternative convention for the 
proton charge radius is being discussed in Ref.~\cite{FrMaSp1997}. 
We refer to this convention as the 
Friar--Martorell--Sprung (FMS) convention as it has been advocated
in Ref.~\cite{FrMaSp1997}.
In FMS conventions [see also Eq.~\eqref{repl}], 
the mean square proton charge radius includes the Darwin--Foldy correction,
\begin{equation}
\label{conv_p}
\left< r^2 \right>_p^\nuc = 
\left< r^2 \right>_p^\ato +
\left< r^2 \right>^{\rm DF}_p \,.
\end{equation}
According to Ref.~\cite{FrMaSp1997},
the alternative mean square charge radius can be written
as the slope of a modified Sachs form factor
\begin{subequations}
\begin{align}
\widetilde G_E(q^2) = & \;
\frac{G_E(q^2)}{ \sqrt{ 1 - q^2 / 4 \, m_p^2} } \,,
\\[0.5ex]
\;
\left< r^2 \right>_p^\nuc =& \; 6 \hbar^2 \left. 
\frac{\partial \widetilde G_E(q^2)}{\partial q^2} \right|_{q^2 = 0}  \,.
\end{align}
\end{subequations}
The authors of Ref.~\cite{FrMaSp1997}
show that this representation
is better adapted to the relativistically invariant representation of the 
Rosenbluth~\cite{Ro1950} formula. 

If one uses the FMS convention~\eqref{conv_p} for the proton charge radius,
then, even for a pointlike nucleus, the nuclear size correction is nonvanishing
for the atomic binding energy. Therefore, the ATP definition of the
charge radii has been favored in Ref.~\cite{EiGrSh2001},  who have argued in
Sec.~7.1.1 of the cited literature reference that in Ref.~\cite{FrMaSp1997}, 
``it is suggested to include the Darwin--Foldy contribution in the definition of the
nuclear charge radius.  While one can use any consistent definition of the
nuclear charge radius, this particular choice seems to us to be unattractive
since in this case even a truly pointlike particle in the sense of quantum
field theory (say an electron) would have a finite charge radius even in
zero-order approximation.'' Here, the authors refer to the zeroth-order
approximation as the one without any radiative corrections. 
In order to ease our mind, we may note that the
Darwin--Foldy correction vanishes for an infinitely heavy, point nucleus,
i.e., in the limit of $m_N \to \infty$. In this limit, the ATP and FMS
conventions for the charge radius definition are in agreement.
Due to the uncertainty principle, one cannot locate
a particle and therefore, its charge, to better than its 
Compton wavelength. The Darwin--Foldy correction is of this magnitude.

\begin{figure*}
\begin{center}
\begin{minipage}{16cm}
\begin{center}
\includegraphics[width=0.7\linewidth]{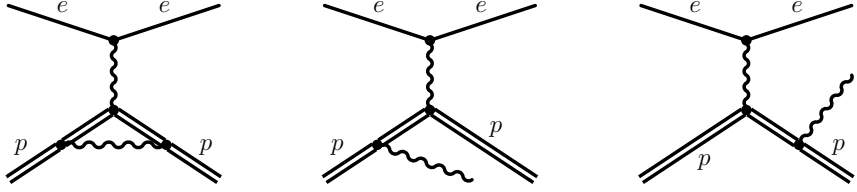}
\caption{\label{fig1} Radiative vertex correction
to electron-proton scattering (heavy line)
and associated bremsstrahlung diagrams.}
\end{center}
\end{minipage}
\end{center}
\end{figure*}

There is thus an immediate need to clarify
which conventions have actually been used in the 
determinations of the proton charge radius 
in Refs.~\cite{Ro2000proton,Si2003proton,Si2007} and in the 
CODATA adjustment in Ref.~\cite{MoTaNe2008}. We observe:

{\em (i)} According to the first (unnumbered) equation
given on p.~410 of Ref.~\cite{Si2007},
one may infer that in Refs.~\cite{Ro2000proton,Si2003proton,Si2007},
the form factor $G_E$ (or merely an infrared safe, subtracted variant of it) 
and not $\widetilde G_E$ is being used for the 
determination of the rms radius given in Eq.~\eqref{rmsproton};
this observation can be confirmed~\cite{Si2010priv}.
This means that the values given in 
Refs.~\cite{Ro2000proton,Si2003proton,Si2007} and 
in particular in Eq.~\eqref{rmsproton} are 
in agreement with the atomic physics conventions mentioned above.
The same applies to the values given in Eq.~\eqref{rpSCAT}.

{\em (ii)} In the latest CODATA adjustment given in
Ref.~\cite{MoTaNe2008}, the conventions are consistent
with those used here and in Ref.~\cite{HuEtAl1998},
although this is somewhat less obvious.
First of all, we reemphasize that, as evident from 
Eq.~(10) of Ref.~\cite{MoTaNe2008},
the atomic binding energy is defined to 
include the Dirac--Foldy term in the CODATA adjustment.
For the deuteron, there is no Darwin--Foldy correction
in the atomic binding energy (see~\cite{PaKa1995}).
In Ref.~\cite{MoTaNe2008}, 
the authors still use the Darwin--Foldy correction
in the atomic part of the energy (even for deuterium) and later add
the Darwin--Foldy correction back on to the deuterium radius.
This is consistently done in all CODATA adjustments
since the 1998 adjustment (see Ref.~\cite{MoTa2000}),
and a pertinent discussion
can be found in Appendix A8 near Eq.~(A56) of Ref.~\cite{MoTa2000}.
A superficial reading of  Appendix~A8
of Ref.~\cite{MoTa2000} might otherwise 
suggest that the Darwin--Foldy correction has been inadvertently
added to the deuteron (rather than proton!) 
nuclear charge radius in Ref.~\cite{MoTa2000};
but this addition is compensated by the inclusion of the 
Darwin-Foldy term even for bound-state deuterium energy levels, 
where according to Ref.~\cite{PaKa1995}, this term should
have been excluded. Therefore, both proton and 
deuteron charge radii given in Ref.~\cite{MoTaNe2008}
correspond to ATP conventions~\cite{HuEtAl1998}
and exclude the Darwin-Foldy term from the nuclear radii
for both proton and deuteron.

%
%
\section{Radiative Corrections}
\label{radcorrec}

We now turn our attention to the role of radiative corrections
to the proton line in both nuclear physics scattering 
experiments as well as in the determination of atomic 
binding energy levels. We recall once 
more that in atomic physics (ATP) conventions, the proton charge radius 
is defined as the slope of a subtracted electric $G_E$ Sachs form factor,
$\langle r^2 \rangle^\ato_p =
6 \hbar^2 \partial G_E/\partial q^2 |_{q^2 = 0}$.
The electric and magnetic $G_E$ and $G_M$ form 
factors for the proton (a spin-$\tfrac12$ particle)
are related to the Dirac $F_1$ and Pauli $F_2$ form 
factors by the relations
\begin{subequations}
\begin{align}
G_E(q^2) =& \; F_1(q^2) + \frac{q^2}{4 (m_p c)^2} \, F_2(q^2) \,,
\\[0.5ex]
G_M(q^2) =& \; F_1(q^2) + F_2(q^2) \,, \qquad
F_2(0) = \varkappa_p \,,
\end{align}
\end{subequations}
where $\varkappa_p = (g_p - 2)/2 = 1.792\,847\,356(23) $ 
gives the anomalous magnetic moment of the proton
(see Ref.~\cite{MoTaNe2008}).
For a point particle like the electron, we have 
$\varkappa_e = \alpha/(2 \pi) \ll 1$, but for 
a particle like the proton,
the bulk of the contribution to 
$\varkappa_p$ is from the internal structure, whereas 
a tiny correction also is due to its electromagnetic nature.

Let us therefore make the following separation,
\begin{subequations}
\label{separation}
\begin{align}
G_E(q^2) =& \; \overline G_E(q^2) + G_E^{\rm QED}(q^2) \,,
\\[0.5ex]
F_1(q^2) =& \; \overline F_1(q^2) + F_1^{\rm QED}(q^2) \,,
\\[0.5ex]
F_2(q^2) =& \; \overline F_2(q^2) + F_2^{\rm QED}(q^2) \,.
\\[0.5ex]
\varkappa_p  =& \; \overline \varkappa_p + \varkappa_p^{\rm QED} \,.
\end{align}
\end{subequations}
Here, the overlined quantities represent the contributions to the 
proton form factor due to its internal structure, 
whereas the quantities with the superscript QED represent the 
contributions due to QED point-particle theory.

The one-loop slope of $G_E^{\rm QED}$ is infrared 
divergent solely due to QED vertex corrections 
(see Fig.~\ref{fig1}), but the slope of 
$\overline G_E$ is infrared safe.
Indeed, the slope of the Dirac $F_1$ form factor 
of any charged spin-$\tfrac12$ particle is infrared 
divergent, and
the coefficient of the logarithmic infrared 
divergence of $F_1$ even
is independent of the nuclear spin~\cite{ItZu1980}. 
The infrared problem persists both in the atomic physics determination
of the nuclear charge radius as well as in the determination
from scattering data. From Chap.~7 of 
Ref.~\cite{ItZu1980}, we know that the 
{\em pure QED} radiative contribution from the 
first diagram in Fig.~\ref{fig1} corresponds to the 
following replacement of the Dirac current $\gamma^\mu$
of the proton,
\begin{equation}
\label{radcor}
\gamma^\mu \to 
\gamma^\mu +
\left\{ \gamma^\mu \, \left[ F_1^{\rm QED}(q^2) - 1\right]+ 
\ii \, \frac{\sigma^{\mu\nu} \, q_\nu}{2 m_p c} 
F_2^{\rm QED}(q^2) \right\} \,,
\end{equation}
where $F_1^{\rm QED}$ and $F_2^{\rm QED}$ are the QED expressions
for the Dirac and Pauli form factors of a spin-$\tfrac12$ point
particle with the proton mass, as given in 
Eqs.~(7-60) and (7-58) of Ref.~\cite{ItZu1980}, respectively.
In particular, we have $F_2^{\rm QED}(0) = \alpha/(2 \pi)$.
From Eqs.~(3.36) and (4.14) of Ref.~\cite{MaTj2000},
and from Eq.~(A76) of Ref.~\cite{VaEtAl2000}, 
we may conclude that the radiative QED term in curly brackets 
in Eq.~\eqref{radcor} is indeed subtracted when the 
radiative corrections to electron-proton scattering
are eliminated from experimental scattering data.

We now have to turn our attention to atomic physics and
identify the nuclear self-energy as the 
correction to atomic energy levels corresponding to the 
terms in curly brackets in Eq.~\eqref{radcor}, which 
are subtracted from scattering data. 
To this end, we first make a slight detour 
and observe that the proton radius 
definition according to 
\begin{align}
\langle r^2 \rangle_p^{\rm ATP} =& \;
6 \hbar^2 \left.  \frac{\partial \overline G_E(q^2)}{\partial q^2} \right|_{q^2 = 0} 
\nonumber\\[0.5ex]
=& \;
6 \hbar^2 \left.  \frac{\partial 
\overline F_1(q^2)}{\partial q^2} \right|_{q^2 = 0} +
\frac{3 \hbar^2}{2 \, (m_p c)^2} \, \overline \varkappa_p \,,
\end{align}
entails a term originating from the ``internal'' contribution
$\overline \varkappa_p$ to the anomalous magnetic moment of the proton.
We note that $\varkappa_p \approx \overline \varkappa_p$
because $\varkappa_p^{\rm QED} = \alpha/(2 \pi) \ll \varkappa_p$ is small.
The anomalous magnetic moment of the 
electron shifts $S$ state energy levels 
[see Eqs.~(42) and~(44) of Ref.~\cite{EiGrSh2001}] by 
\begin{align} 
E_{\varkappa_e} =& \; \left( \frac{\alpha}{2 \pi} \right) \,
\frac{(Z\alpha)^4}{n^3} \, 
\left( \frac{m_r}{m_e} \right)^3 \, m_e c^2 \, 
\delta_{\ell 0} 
\nonumber\\
=& \; \varkappa_e \, \frac{(Z\alpha)^4}{n^3} \, 
\left( \frac{m_r}{m_e} \right)^3 \, m_e c^2 \, 
\delta_{\ell 0}\,.
\end{align}
The corresponding proton line contribution reads
\begin{align} 
\label{ENS1}
E_{\varkappa_p} =& \; 
\overline \varkappa_p \, \frac{(Z\alpha)^4}{n^3} \, 
\left( \frac{m_r}{m_p} \right)^2 \, m_r c^2 \, 
\delta_{\ell 0} 
\nonumber\\
=& \;
\frac23 \left( \frac{m_r}{m_e} \right)^3 \,
\frac{(Z \alpha)^4 m_e c^2}{n^3 \, \lambdabar_C^2} \, 
\left( \frac32 \frac{\hbar^2}{( m_p \, c )^2} \,
\overline\varkappa_p \right) \,
\delta_{\ell 0} \,.
\end{align}
The slope of the $F_1$ form factor of the proton leads to
an energy correction,
\begin{align} 
\label{ENS2}
E_{F_1} =& \; 
\frac23 \left( \frac{m_r}{m_e} \right)^3 \,
\frac{(Z \alpha)^4 m_e c^2}{n^3 \, \lambdabar_C^2} \, 
\left( 6 \hbar^2 \left.  \frac{\partial \overline F_1(q^2)}{\partial q^2} 
\right|_{q^2 = 0} \right) \,
\delta_{\ell 0} \,.
\end{align}
The sum of $E_{\varkappa_p}$ and $E_{F_1}$ is
\begin{align} 
\label{ENS3}
E_{\rm NS} =& \; 
\frac23 \left( \frac{m_r}{m_e} \right)^3 \,
\frac{(Z \alpha)^4 m_e c^2}{n^3 \, \lambdabar_C^2} \, 
\nonumber\\[0.5ex]
& \; \times \left( 6 \hbar^2 \left. \frac{\partial 
\overline F_1(q^2)}{\partial q^2} 
\right|_{q^2 = 0} +  \frac32 \frac{\hbar^2}{( m_p \, c )^2} 
\overline \varkappa_p \right) 
\nonumber\\[0.5ex]
=& \; 
\frac23 \left( \frac{m_r}{m_e} \right)^3 \,
\frac{(Z \alpha)^4 m_e c^2}{n^3 \, \lambdabar_C^2} \, 
\langle r^2 \rangle_p^{\rm ATP} \, \delta_{\ell 0} 
\nonumber\\[0.5ex]
=& \; 
\frac23 \, \hbar c \,
\langle r^2 \rangle_p^{\rm ATP} \, 
\biggl< \pi (Z\alpha) \delta^3(r) \biggr>_{nS} \, 
\delta_{\ell 0} \,,
\end{align}
which corresponds to the well-known expression~\eqref{deltaENS}
and clarifies that indeed, the anomalous magnetic term
of the proton forms part of the nuclear size 
correction in the ATP conventions, and that indeed,
this convention is in agreement with the 
mean-square-radius of the proton being defined as the
slope of the Sachs $G_E$ form factor.
The expectation value of the Dirac $\delta$ for $S$ states is
\begin{equation}
\biggl< \pi (Z\alpha) \delta^3(r) \biggr>_{nS} =
\frac{ (Z\alpha)^3 }{n^3} \left( \frac{m_r \, c}{\hbar} \right)^3 \,.
\end{equation}

In order to describe the two-body interaction
including the nuclear-size effect, we
have to consult the two-body Breit Hamiltonian,
and temporarily switch to natural units with $\hbar = c = \epsilon_0 = 1$.
The Schr\"{o}dinger Hamiltonian of the two-body 
system consisting of an orbiting spin-$\tfrac12$ particle 
of mass $m_e$ (the electron) and a spin-$\tfrac12$ nucleus of 
mass $m_N$ and nuclear charge number $Z$ is
\begin{equation}
H_S = \frac{\vec p^2}{2 m_r} - \frac{Z \alpha}{r} \,,
\end{equation}
where $m_r$ is the reduced mass of the system.
The two-body Breit--Pauli Hamiltonian with
anomalous magnetic moment corrections for electron
and proton reads 
\begin{align} 
\label{HBP}
& H = 
- \frac{\vec p^4}{8 m_e^3} 
- \frac{\vec p^4}{8 m_p^3} 
\nonumber\\[0.5ex]
& \; + \frac{2 Z \alpha}{3} 
\left( \frac{3}{4 m_e^2} + \frac{3}{4 m_p^2} + 
\left< r^2 \right>_p^{\rm ATP} \right) \, \pi \delta^3(r) 
\nonumber\\[0.5ex]
& \; -  \frac{Z\alpha}{2 m_e m_p r} \, 
\vec p \left( \frac{\delta^{ij}}{r} +\frac{r^i \, r^j}{r^3} \right) \vec p
+ \underbrace{(1 + 2 \varkappa_e) \frac{Z\alpha}{4 m_e^2 r^3} 
\vec L \cdot \vec \sigma_e}_{\rm fs}
\nonumber\\[0.5ex]
& \; + 
\underbrace{(1 + \varkappa_e) \frac{Z\alpha}{2 m_e m_p r^3} 
\vec L \cdot \vec \sigma_e}_{\rm fs}
+ \underbrace{(1 + 2\varkappa_p) \frac{Z\alpha}{4 m_p^2 r^3} 
\vec L \cdot \vec\sigma_p}_{\rm hfs}
\nonumber\\[0.5ex]
& \; 
+ \underbrace{(1 + \varkappa_p) \frac{Z\alpha}{2 m_e m_p r^3} 
\vec L \cdot \vec \sigma_p}_{\rm hfs}
+ \underbrace{
\frac{(1 + \varkappa_e) (1 + \varkappa_p)Z \alpha}{4 m_e m_p r^3}}_{\rm hfs}
\nonumber\\[0.5ex]
& \; \times
\underbrace{ \left\{ 
 \frac{8 \pi}{3} \vec\sigma_e \cdot \vec\sigma_p \; \delta^3(r) +
 3 \frac{\vec \sigma_e \cdot \vec r \; \vec \sigma_p \cdot \vec r}{r^5} -
 \frac{\vec \sigma_e \cdot \vec\sigma_p}{r^3} \right\}}_{\rm hfs} \,.
\end{align}
Here, we keep the nuclear charge number
$Z$ as a variable so that the expression below 
can be readily generalized to a spin-$\tfrac12$ nucleus 
with $Z \neq 1$. Terms labeled with ``fs'' are relevant for the fine structure, 
whereas terms labeled by ``hfs'' correspond to the 
hyperfine structure. All terms in Eq.~\eqref{HBP} proportional to the 
electron anomaly $\varkappa_e$ contribute to the fine structure and 
thus, to the Lamb shift of these states. Except for the 
term incorporated into the nuclear mean-square charge radius 
$\left< r^2 \right>_p^{\rm ATP}$, the terms proportional to the 
(complete) 
proton anomalous magnetic moment $\varkappa_p$ listed in Eq.~\eqref{HBP}
influence only the exchange of a magnetic photon of the nucleus and the 
orbiting particle, i.e., the hyperfine structure
(not the Lamb shift). Hyperfine structure effects are 
by definition excluded from the Lamb shift~\cite{SaYe1990}.

We now have to carefully define the nuclear self-energy
for $S$ states so that the QED contributions to the 
proton factors, which were excluded from the form factors 
used in Eq.~\eqref{ENS1},~\eqref{ENS2}, and~\eqref{ENS3}
for the evaluation of the nuclear-size correction,
are included instead into the nuclear self-energy.
Furthermore, we notice that the ``fs'' terms,
which are proportional to the spin-orbit coupling
$\vec L \cdot \vec \sigma_e$ of the electron, 
are defined with the 
phenomenologically inserted anomalous magnetic 
moment $\varkappa_e$ of the electron.
We also have to define the nuclear self-energy so that 
the terms due to the $\varkappa_p^{\rm QED} = F_2^{\rm QED}(0)$ part of
proton anomalous moment, which are already included in the 
phenomenologically inserted full anomalous magnetic 
moment $\varkappa_p$ in the $\vec L \cdot \vec \sigma_p$ terms
in Eq.~\eqref{HBP}, are not 
double counted in the nuclear self-energy.

The $F_1$ form factor of the {\em electron} induces the 
following correction to the Lamb shift,
\begin{equation}
\label{ENSEe}
E_e = 
\frac{\alpha (Z\alpha)^4}{\pi n^3} 
\left( \frac{m_r}{m_e} \right)^3 m_e c^2
\left\{ \frac43 \ln\left( \frac{m_e}{2 \epsilon} \right) 
+ \frac{11}{18} + \frac12 \right\} \delta_{\ell 0}\,,
\end{equation}
where $\epsilon$ is a noncovariant, infrared 
cutoff parameter~\cite{Pa1993,JePa2002}. The 
matching with the covariant photon mass
has given rise to some discussion~\cite{LambFirst}
in the early days of quantum electrodynamics.
The terms $11/18$ and $1/2$ are from the 
nonlogarithmic term of the slope of the $F_1 = F_1^{\rm QED}$
form factor of the {\em electron}, and the term $1/2$ is due to 
$\varkappa_e = \varkappa_e^{\rm QED}$.
From the proton line, we have the following,
corresponding contribution due to the 
QED parts $F_1^{\rm QED}$ and $F_2^{\rm QED}$ of the 
proton form factors,
\begin{equation}
\label{ENSE1}
E_{\mathrm{NSE},1} = 
\frac{Z (Z\alpha)^5}{\pi \, n^3} 
\left( \frac{m_r}{m_p} \right)^2 \, m_r c^2 \,
\left\{ \frac43 \ln\left( \frac{m_p}{2 \epsilon} \right) 
+\frac{10}{9}  \right\} \delta_{\ell 0},
\end{equation}
for $S$ states.
The infrared divergence in~\eqref{ENSE1}, which for scattering
experiments is compensated by bremsstrahlung diagrams, is cut off 
in atomic physics at the binding energy scale $(Z\alpha)^2 m_r$.
The matching low-energy contribution~\cite{Be1947} contains the Bethe logarithm 
$\ln k_0(n,\ell)$,
\begin{align}
\label{ENSE2}
& E_{\mathrm{NSE},2} = 
\frac{Z (Z\alpha)^5}{\pi \, n^3} \;
\left( \frac{m_r}{m_p} \right)^2 \, m_r c^2 
\\[0.5ex]
& \; \times 
\left\{ \frac43 \ln\left( \frac{2 \epsilon}{(Z\alpha)^2 \, m_r} \right) \,
\delta_{\ell 0} - \frac43 \ln k_0(n,\ell) \right\} \,.
\nonumber
\end{align}
The sum of~\eqref{ENSE1} and~\eqref{ENSE2} is free from the 
scale separation parameter $\epsilon$ and reads
\begin{align}
\label{ENSE}
& E_{\mathrm{NSE}} = \;
\frac{Z (Z\alpha)^5}{\pi \, n^3} \;
\left( \frac{m_r}{m_p} \right)^2 \, m_r c^2 \, 
\\[0.5ex]
& \; \times \left\{ \left[ \frac43 \ln\left( \frac{m_p}{m_r \, (Z\alpha)^2} \right) \,
+\frac{10}{9} \right] \, \delta_{\ell 0}
- \frac43 \ln k_0(n,\ell) \right\} \,.
\nonumber
\end{align}
This expression generalizes the nuclear self-energy 
to states with nonvanishing angular momenta.

A few remarks are in order. First, we observe that the 
anomalous magnetic moment $\varkappa_e$,
due to the spin-orbit coupling of the electron, 
leads to the following term in the self-energy
for non-$S$ states, 
\begin{equation}
E' = \frac{\alpha (Z\alpha)^4}{\pi \, n^3} \;
\left( \frac{m_r}{m_e} \right)^2 \, m_e c^2 \, 
\left( - \frac{1}{2 \kappa (2 \ell + 1)} \right) \,
\left( 1 - \delta_{\ell 0} \right) \,,
\end{equation}
where $\kappa = 2 (\ell-j) \, (j + 1/2)$ is the 
Dirac quantum number.
The analogous term due to the proton line is
given in Eq.~(153) of Ref.~\cite{EiGrSh2001},
\begin{equation}
E'' = \frac{Z (Z\alpha)^5}{\pi \, n^3} \;
\frac{m_r}{m_p} \, m_r c^2 \, 
\left( - \frac{1}{2 \kappa (2 \ell + 1)} \right) \,
\left( 1 - \delta_{\ell 0} \right) \,.
\end{equation}
However, this term originates from the spin-orbit
coupling of the proton and is already contained in the 
phenomenologically inserted $\varkappa_p$ in Eq.~\eqref{HBP}.
We therefore exclude this term from the definition
of the nuclear self-energy for non-$S$ states.

The Bethe logarithm term $\ln k_0(n,\ell)$ 
in Eq.~\eqref{ENSE}, 
for a general hydrogenic state,
can be derived on the basis of nonrelativistic 
perturbation theory for the two-body system~\cite{Pa1998}.  
It is a consequence of the low-energy
part of the proton self energy, which is due to the exchange of
low-energy photons along the proton line.
The logarithmic term (in $\epsilon$) in the low-energy part
given in Eq.~\eqref{ENSE2} compensates the infrared divergence of the 
$F_1^{\rm QED}$ form factors at the atomic binding energy scale,
in analogy to the cancellation of infrared divergences in the
vertex and bremsstrahlung diagrams (for scattering).
The definition~\eqref{ENSE} takes into account the quantum
electrodynamic properties of the proton 
(as a charged spin-$\tfrac12$ particle) and is compatible with
the radiative corrections listed in Eq.~\eqref{radcor}. 
The nonlogarithmic term $10/9$ in Eq.~\eqref{ENSE},
which constitutes an addition to 
expression for the nuclear self-energy given in Ref.~\cite{Pa1995radrec},
corresponds to a shift of the proton
radius by about $\delta r_p = 0.0001 \, \fm $ or alternatively 
to 0.20~kHz in frequency
units for the hydrogen-deuterium isotope shift of the $1S$--$2S$ transition.

Finally, one might argue with respect to our result~\eqref{ENSE} 
that the starting point of QED calculations connected with the
proton is a proton with form factors that are due to strong interactions,
and that, therefore,
all QED integrals are cut off from above by the proton radius,
i.e., that the proton structure significantly influences the
exchange of virtual photons of wavelengths shorter than
its own size.
However, the $\epsilon$ parameter is an infrared cutoff,
and it stems from the infrared divergence of the 
$F_1^{\rm QED}$ form factor of the proton. Indeed, the physically
appropriate values for the $\epsilon$ parameter are in the region
\begin{equation}
\label{restriction}
(Z \alpha)^2 m_e \ll \epsilon \ll m_e \ll \Lambda, m_p \,,
\end{equation}
where $\Lambda = \sqrt{0.71 {\rm GeV}^2} = 0.842 {\rm GeV}$ 
is a parameter characterizing the proton size.
The first inequality in Eq.~\eqref{restriction}
is due to the requirement that all bound-state poles 
must be integrated over in evaluating the
low-energy part of the self-energy, and the latter
inequality is due to the relativistic nature of the
electron mass scale (separation from the high-energy part).
The coefficient multiplying the logarithmic infrared divergence
in Eq.~\eqref{ENSE1} is the same whether we use
$\ln[ \Lambda/(2 \epsilon) ]$
for the logarithmic term,
as done in Eq.~(154) of Ref.~\cite{EiGrSh2001}, or
$\ln[ m_p/(2 \epsilon) ]$,
with $m_p = 0.938 \, {\rm GeV}$, as done here. The proton structure thus
does not influence the leading coefficient of the
logarithmic divergence of the infrared behavior of the
form factor. The question then is which portion of the
proton structure should be counted as a radiative nuclear
self energy, and which should be counted as a proton structure
correction. Here, we advocate the view that if the
radiative corrections to scattering are accounted for
by the $F_1^{\rm QED}$ and $F_2^{\rm QED}$ contributions to the 
form factors of the
proton (treated as a point-like QED particle), then, also,
the proton should be treated as a point-like QED particle
in evaluating the nuclear self energy in atomic physics.
The difference of the above mentioned two logarithms,
\begin{equation}
\ln\left( \frac{\Lambda}{2 \epsilon} \right) -
\ln\left( \frac{m_p}{2 \epsilon} \right) =
\ln\left( \frac{\Lambda}{m_p} \right) \,,
\end{equation}
is infrared safe, part of $\overline F_1$
and therefore, part of the nuclear-size correction
in our conventions.
While this view is at variance with the result obtained in
Eq.~(154) Ref.~\cite{EiGrSh2001}, we stress that the precise definition
of the nuclear self energy depends on the way in which
radiative corrections to the proton line are subtracted
in atomic physics and in scattering experiments. Our view
thus holds relative to the conventions used for
subtracting radiative corrections to the proton line in scattering
experiments.

In the evaluation of higher-order nuclear structure corrections to 
the bound-state spectrum,
we only have to remember that the $G_E$ inferred from
scattering experiments is not the full $G_E$ of the 
proton, but with the 
contributions from $F_1^{\rm QED}$ and $F_2^{\rm QED}$ subtracted.
The QED contributions to the proton form factors 
thus have to be treated separately.
One example can immediately be given as follows.
It is known~\cite{Ze1956} 
that the third Zemach moment correction~\cite{FrSi2005} 
to bound-state energy levels is obtained
in second order perturbation theory involving the 
correction to the Coulomb potential due to the 
finite size of the nucleus, or in other words, 
in the second order with respect to the 
slope of the proton form factor. 
The contribution from the radiative correction due to the 
QED contribution
$F_1^{\rm QED}(q^2)$ to the Dirac form factor of the 
proton is thus neglected in the derivation of the Zemach 
correction. If desired, this effect can be added back on.
It is easy to write down the  leading logarithm
\begin{align}
E^{(2)}(nS) = & \;
- \frac{8}{27} \,
\frac{\alpha^2 (Z\alpha)^6}{\pi^2 \, n^3} 
\left( \frac{m_r^2}{m_p^2} \right)^2 \, m_r c^2
\nonumber\\[0.5ex]
& \; \times \ln^2\left( \frac{m_p}{m_r \, (Z\alpha)^2} \right) \,
\ln\left[ (Z\alpha)^{-2} \right] \,
\end{align}
for the energy correction to $S$ states 
in the second order of perturbation theory 
with respect to the proton self-energy.
The numerical value of this correction is tiny because it is 
of high order in the electron-proton mass ratio.

%
%
\section{Conclusions}
\label{conclu}

In this paper, we have analyzed the role of the Darwin-Foldy term in the
determination of atomic energy levels and in nuclear physics scattering
experiments that measure the proton form factors.  In Sec.~\ref{dfterm}, we
have clarified that all current determinations of the proton charge radius,
both in atomic as well as in nuclear physics, are based on the slope of the
electric Sachs $G_E$ form factor~\eqref{rATP} and thus exclude the Darwin-Foldy
term.  This applies to both the value given in Eq.~\eqref{rmsproton} as well as
the value given in Eq.~\eqref{rmspCODATA}.  The latter value is determined on
the basis of a comparison of the most accurately measured transition
frequencies in atomic hydrogen to theory~\cite{JeKoLBMoTa2005}.

Furthermore, we have clarified that although a superficial reading of
Appendix~A8 of Ref.~\cite{MoTa2000} might suggest that the
Darwin--Foldy correction has been inadvertently added to the deuteron
(rather than proton!) nuclear charge radius in Ref.~\cite{MoTa2000}, this is
actually not the case: the Darwin--Foldy correction is absent for
deuterium~\cite{PaKa1995}, and if one includes this contribution
into atomic energy levels, even for deuterium, as done by the authors of
Refs.~\cite{MoTa2000,MoTaNe2008}, then one has to add the Darwin--Foldy
correction back on to the deuteron radius as is done in 
Appendix~A8 of Ref.~\cite{MoTa2000} (as well as, consistently, in all CODATA
adjustments since 1998).

These observations imply that the Darwin-Foldy term is 
excluded from both values of the proton radii given in 
Eqs.~\eqref{rmsproton} and~\eqref{rmspCODATA},
which in addition are in good mutual numerical agreement.
In Ref.~\cite{PoEtAl2010}, the proton radius is
given as $r_p = 0.84184(67) \, {\rm{fm}}$,
in disagreement with Eqs.~\eqref{rmsproton} and~\eqref{rmspCODATA},
and with the Darwin-Foldy term likewise being excluded.
A conceivable accidental incompatibility of the 
conventions used in 
Refs.~\cite{Ro2000proton,Si2003proton,Si2007,MoTaNe2008,PoEtAl2010} for the proton radius 
therefore cannot be the reason for the observed discrepancy.

In Sec.~\ref{radcorrec}, we turn our attention to radiative corrections to
electron-proton scattering.  Indeed, the infrared divergence of the Sachs $G_E$
form factor at the one-loop level enters both the atomic physics determination
of the proton radius as well as the nuclear physics determination from,
e.g., scattering data. While it is clearly stated in
Refs.~\cite{Ro2000proton,Si2003proton,Si2007} that the proton charge radius is
determined based on scattering data for $G_E$,
it is useful to remark that the slope of the $G_E$ form
factor is determined in each case from {\em subtracted} scattering data: namely, the
infrared divergent, QED radiative corrections whose exact values depend on the
acceptance range of the detectors are subtracted when the slope of the form
factor is being calculated. Theoretical input data for the radiative
corrections is available from a number of sources, notably,
Refs.~\cite{MoTs1969,MaTj2000,VaEtAl2000,EnEtAl2001,BlMeTj2005}.
Equation~\eqref{ENSE} represents an attempt to define the nuclear self energy
in full compatibility with the subtractions of radiative corrections to
electron-proton scattering, notably, Eqs.~(3.36) and (4.14) of
Ref.~\cite{MaTj2000}, and Eq.~(A76) of Ref.~\cite{VaEtAl2000}. The 
result~\eqref{ENSE} also represents a generalization of the nuclear self energy 
to non-$S$ states ($\ell \neq 0$).

Finally, let us remember that the 
nuclear self energy is a numerically rather tiny 
correction, especially for systems with a small mass ratio of the 
orbiting particle versus the nucleus.
Likewise, the consistent subtraction of the 
radiative corrections listed in Eq.~\eqref{radcor}
in a scattering experiment depends on 
the availability of accurate data at low momentum transfer,
and on the accurate subtraction of other contributions, such 
as double scattering (including effects due to the polarizability 
of the proton) and double bremsstrahlung. 
It is known~\cite{MoTs1969,MaTj2000,VaEtAl2000,EnEtAl2001,BlMeTj2005} that the 
contribution of the latter effects may be numerically more
significant than that of the QED terms in curly brackets in
Eq.~\eqref{radcor}.
Currently, the uncertainty induced by the nonlogarithmic 
terms is negligible on the scale of the total uncertainty of the 
proton radius, but still, it is an important conceptual issue
to carry out the analysis of radiative corrections to the 
proton line consistently in atomic and nuclear physics experiments.

\section*{Acknowledgments}

The author acknowledges insightful discussions with J. Friar.
Further helpful conversations with P. J. Mohr, 
M. Vanderhaeghen, C. G. Parthey, I.  Sick, R.
Pohl, N. Kolachevsky, Th.~Udem, and T.~W.~H\"{a}nsch are gratefully
acknowledged.  This research has been 
supported by the National Science
Foundation and by the National Institute of Standards
and Technology (Precision Measurement Grant).

\end{document}